\documentclass[10pt,letterpaper,compsoc,conference]{iiswc26}

\usepackage{cite}
\usepackage{amsmath,amssymb,amsfonts}
\usepackage{algorithmic}
\usepackage{graphicx}
\usepackage[dvipsnames]{xcolor}
\usepackage[final]{microtype}
\usepackage[italic]{mathastext}
\usepackage{libertine}
\usepackage[T1]{fontenc}
\usepackage{textcomp}
\usepackage[varqu,varl]{zi4}
\usepackage[all]{nowidow}
\usepackage[keeplastbox]{flushend}
\usepackage{fancyhdr}

\usepackage{booktabs}
\usepackage{colortbl}
\usepackage{url}
\usepackage{hyperref}

\graphicspath{{figures/}}

\title{PrefixBench-H100: Characterizing Prefix Reuse and \\Time-to-First-Token in H100 LLM Serving}

\author{
  \IEEEauthorblockN{Omkar Shewale}
  \IEEEauthorblockA{Illinois Institute of Technology\\
                    oshewale@hawk.illinoistech.edu}
  \and
  \IEEEauthorblockN{Deepak Kumar}
  \IEEEauthorblockA{Illinois Institute of Technology\\
                    dkumar15@hawk.illinoistech.edu}
  \and
  \IEEEauthorblockN{Divakar Kumar Yadav}
  \IEEEauthorblockA{University of Wisconsin--Milwaukee\\
                    dkyadav@uwm.edu}
}

\begin{document}
\maketitle

\begin{abstract}
Repeated prompt prefixes are increasingly common in LLM serving
workloads, appearing in system prompts, templated retrieval-augmented
generation pipelines, agent frameworks, and multi-turn conversations.
Modern inference runtimes such as vLLM and TensorRT-LLM provide
mechanisms for reusing previously computed KV-cache state across
requests, yet it remains unclear when prefix reuse materially improves
serving performance on contemporary accelerators and when its benefits
are limited by scheduling, cache granularity, concurrency, or memory
pressure.

This paper presents \textsc{PrefixBench-H100}, a reproducible benchmark
and measurement framework for characterizing prefix reuse on a single
NVIDIA H100. \textsc{PrefixBench-H100} combines controlled synthetic
traces with chat-style and retrieval-style workloads, and evaluates two
widely used LLM serving runtimes under matched workload conditions. The
benchmark varies shared-prefix length, suffix diversity, request
arrival pattern, concurrency, output length, and cache configuration,
while collecting time-to-first-token, inter-token latency, end-to-end
latency, throughput, cache-hit statistics, GPU memory usage, and
selected profiling traces.

The goal of \textsc{PrefixBench-H100} is not to introduce a new caching
algorithm, but to expose the practical operating envelope of prefix
reuse for H100-class LLM serving. The study identifies the regime where
prefix reuse provides substantial first-token latency reductions and
the regime where cache pressure erodes them, while showing that cache
effectiveness itself is largely insensitive to concurrency and output
length; the cross-runtime differences that remain arise above the
cache, in the scheduling layer. By organizing these results around workload
properties rather than runtime-specific tuning anecdotes,
\textsc{PrefixBench-H100} provides actionable guidance for researchers
and practitioners deciding when prefix reuse should be prioritized
over other serving optimizations such as scheduler tuning or chunked
prefill. The benchmark artifact includes workload generators,
measurement scripts, and plotting templates intended to support
reproducible characterization of LLM serving workloads on Hopper-class
GPUs.
\end{abstract}

\section{Introduction}
\label{sec:intro}

The mechanisms underneath modern LLM serving are increasingly
out-pacing their published characterizations. Prefix
caching---introduced by PagedAttention~\cite{kwon2023pagedattention}
under the assumption that KV memory was a scarce resource on
A100-class hardware---is now shipped as a default-on feature in both
vLLM~0.21 and TensorRT-LLM~1.2.1 on the H100~NVL, where 94\,GB of
HBM3 changes that assumption. The question this paper asks is not
\emph{whether} prefix reuse works on Hopper-class hardware, but
\emph{where the bottleneck has moved} now that the cache layer is
no longer scarce.

Our central empirical finding is that on H100~NVL, \emph{whenever
the prefix working set fits within KV capacity}, the dominant
cross-runtime difference has migrated to the \emph{scheduler}
above the cache. We demonstrate this in three ways. First, both runtimes cache
$\approx$\,92\% of the input at an 8192-token shared prefix and
deliver a 5--6.5$\times$ TTFT reduction over a reuse-disabled
baseline, but the absolute TTFT gap between them is essentially
flat across the entire prefix-length sweep
(Section~\ref{sec:seq}). Second, when we move from sequential to
bursty arrival at concurrency~32, the leader on p50 TTFT
\emph{flips} from vLLM to TensorRT-LLM by a factor of 2.4$\times$
(Section~\ref{sec:burst})---not because of any change in cache
effectiveness, but because of a scheduling-layer difference under
load. Third, when we move from burst to fixed-rate arrival at 16\,QPS,
the leader flips back to vLLM by a factor of 9.2$\times$ on
\emph{p95} TTFT (Section~\ref{sec:rag}). Three workloads, two
runtimes, two flips---all attributable to the layer above the
cache. That condition is not universal: once the working set
exceeds KV capacity the cache binds again, for both runtimes at
once (Section~\ref{sec:pressure}).

To establish this finding, we introduce \textsc{PrefixBench-H100},
a reproducible benchmark and measurement framework that drives
vLLM and TensorRT-LLM through identical OpenAI-compatible workloads
on a single H100~NVL. The benchmark maps the operating envelope
along five axes (prefix length, suffix diversity, concurrency,
output length, cache configuration) and uses both synthetic
repeated-prefix traces and realistic RAG-style template traces.
A controlled reuse-on/reuse-off comparison anchors the absolute
cost of disabling prefix caching; a \texttt{tokens\_per\_block}
ablation tests whether the documented memory-vs-kernel tradeoff
remains a useful tuning knob in the TensorRT-LLM~1.x PyTorch
backend (Section~\ref{sec:tpb}).

The contributions of this paper are:

\begin{itemize}
  \item A reusable, open-artifact benchmark and measurement
        framework for prefix-reuse characterization on a single
        H100~NVL, driving both runtimes through one OpenAI-API
        load generator over the same JSONL workloads
        (Section~\ref{sec:method}).
  \item A quantitative reuse-on/reuse-off baseline establishing
        the 5--6.5$\times$ TTFT speedup that prefix reuse provides
        for 7B-class workloads on Hopper, and an empirical map of
        where that speedup is preserved or eroded
        (Section~\ref{sec:seq}--\ref{sec:rag}).
  \item Four load-regime findings specific to the Hopper era:
        (i) cache hit rate is concurrency-insensitive at the
        scale we tested, (ii) the runtime that minimizes p50 TTFT
        under burst is not the runtime that minimizes p95 TTFT
        under fixed-rate arrival, (iii) the documented
        \texttt{tokens\_per\_block} tradeoff in TensorRT-LLM is no
        longer detectable in the PyTorch backend on H100, and
        (iv) eviction onset tracks the working-set-to-capacity
        ratio rather than model size, with cache parity holding
        inside it (Sections~\ref{sec:pressure},
        \ref{sec:discussion}).
  \item A workload-oriented tuning recommendation table for
        operators of H100~LLM serving stacks
        (Table~\ref{tab:tuning}).
\end{itemize}

We treat both runtimes as black boxes used at their
default-recommended configurations and study what an operator can
actually observe. The work is characterization, not system
building. The H100~NVL is not incidental: its 94\,GB of HBM3
raises, rather than removes, the threshold at which prefix-cache
scarcity binds. At 7\,B that threshold sits above any workload we
could construct, so attention belongs one layer up the stack; at
32\,B it is reached within a single GPU
(Section~\ref{sec:pressure}). The governing quantity is the ratio
of prefix working set to KV capacity, and we characterize both
sides of it.

\section{Background}
\label{sec:background}

\subsection{Prefill, decode, and the KV cache}

Autoregressive transformer inference proceeds in two phases. During
\emph{prefill}, the model processes all input tokens in parallel and
writes their key/value activations into a per-layer KV cache. During
\emph{decode}, the model generates output tokens one at a time,
reading the entire prior KV cache at each step. Prefill is
compute-bound and dominates first-token latency for long prompts;
decode is memory-bandwidth bound and dominates per-token latency for
long generations. Cached prefill state is the largest avoidable cost
for workloads where many requests share a common prompt prefix.

\subsection{Prefix caching in vLLM}

vLLM implements automatic prefix caching on top of its PagedAttention
KV-cache layout~\cite{kwon2023pagedattention}, which stores the
per-token cache in fixed-size blocks. When a new request arrives,
vLLM hashes successive blocks of the input and reuses any blocks whose
hashes match a cached entry. The user enables this with the
\texttt{--enable-prefix-caching} server flag (in
vLLM~$\geq$\,0.4 the feature can also be turned on per request).
Eviction is LRU at the block level; block size is configurable but
defaults to 16 tokens.

\subsection{KV-cache reuse in TensorRT-LLM}

TensorRT-LLM exposes an analogous mechanism configured via the
\texttt{KvCacheConfig} dataclass. The two parameters relevant to this
paper are \texttt{enable\_block\_reuse} (a boolean turning reuse on)
and \texttt{tokens\_per\_block} (block granularity, default~32).
The TensorRT-LLM manual notes a tradeoff: smaller blocks unlock more
fine-grained reuse but reduce attention-kernel efficiency, while
larger blocks improve kernel utilization but coarsen the matching
granularity. The manual further notes that reuse only becomes
available once the originating request has completed its prefill,
which has consequences for bursty arrivals (Section~\ref{sec:burst}).

\subsection{The H100 and Hopper-class memory hierarchy}

The H100~NVL provides 94\,GB of HBM3 at $\approx$\,3.9\,TB/s of
bandwidth and fourth-generation Tensor Cores with FP8
support~\cite{nvidia_h100_datasheet}. The large HBM capacity
substantially reduces KV-eviction pressure compared to A100 and
gives the runtime room to retain long shared prefixes even under
moderate concurrency; this paper restricts itself to BF16 weights
and BF16 KV cache to keep precision out of scope.

\section{Methodology}
\label{sec:method}

The artifact for this paper---comprising the
benchmark code, raw per-cell measurement results (102 JSON files
and a merged CSV), all generated figures, and a single Jupyter
notebook (\texttt{analysis.ipynb}) that reproduces every figure
without GPU access---is publicly available at
\url{https://doi.org/10.5281/zenodo.21725505}. A full reproduction
on a single H100~NVL takes approximately three GPU-hours.

\subsection{Hardware and software environment}

All measurements in this paper are collected on a single NVIDIA H100
NVL GPU with 94\,GB of HBM3 memory, driver version 580.105.08, and the
GPU's stock 400\,W power limit. The host is an Azure Ubuntu 24.04 VM.
Persistence mode is enabled to remove first-use driver overhead from
the measurements.

We evaluate two production LLM serving runtimes, used as shipped
without source modifications: vLLM~0.21.0 with PyTorch 2.11.0
(CUDA~13.0) and TensorRT-LLM~1.2.1 with PyTorch 2.9.1 (CUDA~12.8). To
avoid Python dependency conflicts between the two runtimes, each is
installed in an isolated virtual environment. For TensorRT-LLM we use
the PyTorch backend, which is the documented default for serving in
the 1.x release line and exposes the same \texttt{KvCacheConfig}
parameters (including \texttt{enable\_block\_reuse} and
\texttt{tokens\_per\_block}) as the TensorRT backend.

\subsection{Model}

All experiments use \texttt{Qwen/Qwen2-7B-Instruct} (Apache~2.0
license, 7.6\,B parameters, BF16 weights, 32K-token context window).
Qwen2-7B is supported by both runtimes and is small enough to leave
ample HBM headroom on the H100~NVL for KV-cache experiments while
remaining representative of the popular ``7--8\,B Instruct'' class
deployed in production RAG and agent pipelines. We hold the model
fixed across all sweeps; a cross-architecture study is left as
future work.

\subsection{Workloads}
\label{sec:workloads}

\textsc{PrefixBench-H100} drives both runtimes from the same JSONL
workload files through their OpenAI-compatible \texttt{/v1/completions}
endpoints, ensuring that vLLM and TensorRT-LLM see identical prompts
and request schedules. We use two workload generators.

\paragraph{Synthetic repeated-prefix traces.} A controlled generator
emits requests of the form $\langle \text{shared prefix} \rangle
\circ \langle \text{unique suffix} \rangle$, where both segments are
constructed to tokenize to exact token counts under the Qwen2
tokenizer. The generator exposes the shared-prefix length, suffix
length, output length, number of distinct prefixes, and number of
requests as independent knobs. This isolates prefix reuse as the only
source of variation across sweeps.

\paragraph{Realistic RAG-style template traces.} A second generator
emits prompts that resemble retrieval-augmented serving: a long
system prompt, one of eight retrieved documents drawn uniformly per
request, and one of eight query templates. This stresses prefix reuse
under realistic suffix diversity and is used to validate that
synthetic findings transfer.

\subsection{Load driver}
\label{sec:driver}

A single asynchronous Python load driver (\texttt{load\_gen.py},
$\approx$\,300 lines, \texttt{aiohttp}-based) is used for both
runtimes. The driver supports burst (all-at-once) and fixed-rate
inter-arrival patterns. Fixed-rate arrivals are evenly spaced, with
request $i$ released at $i/\mathrm{QPS}$; the driver does not sample
exponential inter-arrival gaps, so the steady-load cells below
characterise deterministic rather than Poisson traffic
(Section~\ref{sec:limits}). It records per-request time-to-first-token
(measured at the first non-empty streamed token), end-to-end latency,
output token count, and any HTTP errors. Per-runtime
\texttt{/metrics} endpoints are scraped before and after each cell so
that per-cell deltas of prefix-cache queries, hits, and reused KV
blocks are recoverable.

\subsection{Sweeps}
\label{sec:sweeps}

We organize the measurement matrix into four sweeps, each emphasizing
a distinct axis. Sweeps are run independently in random order to
mitigate residual cache contamination across cells.

\begin{enumerate}
  \item \textbf{Sequential reuse} (Section~\ref{sec:seq}):
        concurrency~1 with shared-prefix length in
        $\{0,512,2048,4096,8192\}$ tokens. Establishes the
        upper-bound TTFT benefit per runtime.
  \item \textbf{Burst concurrency} (Section~\ref{sec:burst}):
        prefix length fixed at 4096 tokens, in-flight requests in
        $\{1,4,8,16,32\}$. Reveals when concurrency overlaps the
        first-request prefill and suppresses reuse.
  \item \textbf{Block-size ablation} (Section~\ref{sec:tpb}):
        \texttt{tokens\_per\_block} $\in \{32,64,128\}$ for
        TensorRT-LLM. Quantifies the reuse-granularity vs.\
        kernel-efficiency tradeoff documented in the TensorRT-LLM
        manual.
  \item \textbf{Realistic RAG} (Section~\ref{sec:rag}):
        Fixed-rate arrivals at $\{1,4,8,16\}$\,QPS using the
        template-based generator. Confirms whether synthetic
        findings transfer to a workload that more closely resembles
        production traffic.
\end{enumerate}

\subsection{Metrics}

For each cell we report TTFT (p50, p95), end-to-end latency
(p50, p95), output throughput (tokens/s), request throughput
(req/s), prefix-cache hit rate (where the runtime exposes it), and
peak HBM usage scraped from the runtime's Prometheus endpoint. A
small Nsight Systems trace is collected for a representative hit
case and miss case for the case-study figure.

\begin{figure}[t]
  \centering
  \includegraphics[width=0.82\columnwidth]{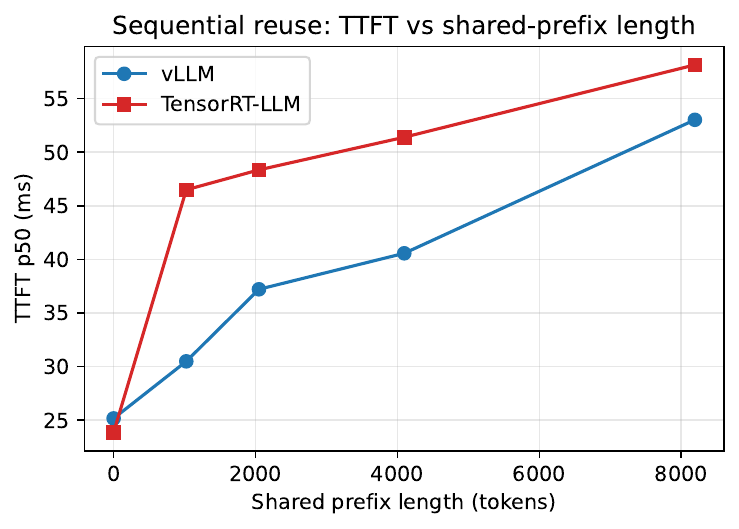}
  \caption{\textbf{Sequential reuse delivers a 5--6.5$\times$ TTFT
           reduction at long prefixes.} p50 TTFT vs.\ shared-prefix
           length at concurrency~1. Both runtimes track each other
           closely; the cross-runtime gap is dominated by
           $\approx$\,8\,ms of per-request overhead, not cache
           behavior. Reuse-off baselines reach 345\,ms (vLLM) and
           306\,ms (TensorRT-LLM) at pl=8192 (Table~\ref{tab:reuse-speedup}).}
  \label{fig:ttft-prefix}
\end{figure}

\section{Results}
\label{sec:results}

\subsection{Sequential reuse: how large is the TTFT speedup, and is it the same in both runtimes?}
\label{sec:seq}

We first establish the upper-bound benefit of prefix reuse by sending
requests one at a time (concurrency~1) across shared-prefix lengths
in $\{0,1024,2048,4096,8192\}$ tokens. The first request in each cell
populates the cache; the median TTFT we report is therefore the
\emph{warm-cache} cost. To anchor what reuse is actually saving, we
re-run the no-shared-prefix (pl=0), 2048, and 8192 cells with
reuse disabled (\texttt{--no-enable-prefix-caching} on vLLM and
\texttt{enable\_block\_reuse: false} on TensorRT-LLM); the
resulting TTFT pairs are reported in Table~\ref{tab:reuse-speedup}.

\begin{table}[t]
  \centering
  \caption{Reuse-on vs.\ reuse-off TTFT (p50, milliseconds) at
           concurrency~1, single shared prefix, 40 requests per
           cell. The speedup column reports the ratio of
           reuse-off~/~reuse-on TTFT.}
  \label{tab:reuse-speedup}
  \footnotesize
  \begin{tabular}{lcccc}
    \toprule
    Runtime & Prefix len & Reuse off & Reuse on & Speedup \\
    \midrule
    vLLM           & 0    & 25.1 & 25.2 & 1.00$\times$ \\
    vLLM           & 2048 & 92.1 & 37.2 & 2.48$\times$ \\
    vLLM           & 8192 & 345.1 & 53.0 & \textbf{6.51$\times$} \\
    \midrule
    TensorRT-LLM   & 0    & 23.7 & 23.9 & 0.99$\times$ \\
    TensorRT-LLM   & 2048 & 90.5 & 48.4 & 1.87$\times$ \\
    TensorRT-LLM   & 8192 & 306.1 & 58.2 & \textbf{5.26$\times$} \\
    \bottomrule
  \end{tabular}
\end{table}

Three observations follow.

\paragraph{Reuse delivers a 5--6.5$\times$ TTFT reduction at an
8192-token shared prefix.} This is the central magnitude reviewers
should remember: turning on prefix reuse reduces p50 first-token
latency by an order of magnitude for the long-prefix regime that
RAG and agent workloads spend most of their time in. The reuse-off
TTFT scales near-linearly with prefix length (the prefill cost
itself), while reuse-on TTFT is nearly flat. Reuse has converted
a prefill cost into a cache lookup.

\paragraph{Cross-runtime cache effectiveness is statistically
identical.} Both runtimes cache 65\%, 78\%, 87\%, and 92\% of the
input at prefix lengths 1024, 2048, 4096, and 8192 tokens
respectively (vLLM hit rate is computed from the
\texttt{prefix\_cache\_hits\_total}~/
\texttt{prefix\_cache\_queries\_total} delta on its
\texttt{/metrics} endpoint; TensorRT-LLM from per-request
\texttt{usage.prompt\_tokens\_details.cached\_tokens}). The cache
mechanism itself, designed in two different codebases, achieves
the same effect.

\paragraph{The cross-runtime gap is per-request overhead, not cache
behavior.} Figure~\ref{fig:ttft-prefix} shows vLLM's curve sitting
$\approx$\,8\,ms below TensorRT-LLM's across the entire sweep---a
gap that exists even at pl=0 where there is no shared prefix to
cache. The gap is therefore upstream of the cache: it reflects the
fixed per-request cost that each runtime imposes regardless of
prefix-reuse activity. Operators should not infer that vLLM is
``better at reuse''---it is roughly equivalent in cache
effectiveness, but its serving stack imposes less first-token
overhead on H100~NVL in the configurations we tested.

\subsection{Burst concurrency: when does reuse survive load?}
\label{sec:burst}

\begin{figure}[t]
  \centering
  \includegraphics[width=0.82\columnwidth]{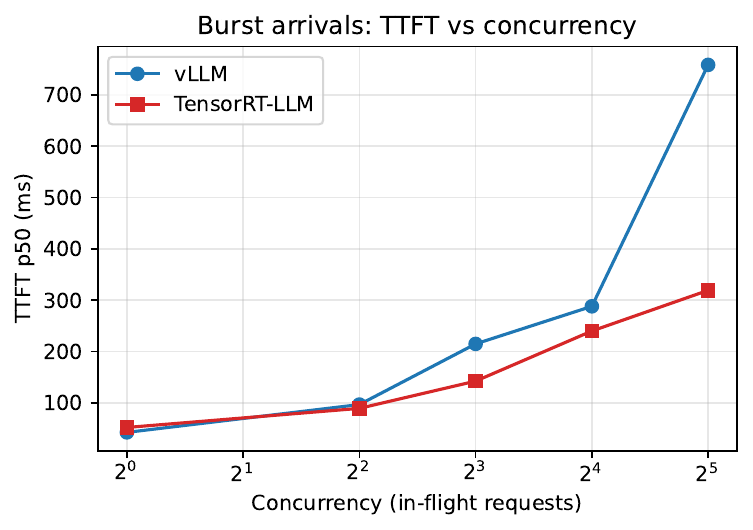}
  \caption{\textbf{TTFT leadership flips between vLLM and
           TensorRT-LLM at c=8.} p50 TTFT vs.\ in-flight
           concurrency at 4096-token shared prefix, burst arrival.
           vLLM leads at c$\leq$4; at c$\geq$16, TensorRT-LLM scales
           linearly while vLLM inflates super-linearly. At c=32,
           TensorRT-LLM has 2.4$\times$ lower p50 TTFT.}
  \label{fig:ttft-concurrency}
\end{figure}

The intuition often cited in the TensorRT-LLM manual~\cite{trtllm_docs}
is that prefix reuse should \emph{degrade} under bursty arrival,
because cache entries only become available after the originating
request completes prefill. To probe this directly, we fix the shared
prefix at 4096 tokens and the number of distinct prefixes at four
(so a burst of 32 in-flight requests draws from four cacheable
prefixes), and sweep in-flight concurrency over
$\{1,4,8,16,32\}$ with 100 requests per cell.
Figure~\ref{fig:ttft-concurrency} plots the resulting p50 TTFT.

Two findings emerge.

\paragraph{TTFT scales much worse for vLLM than for TensorRT-LLM under
load.} At concurrency~1, vLLM has the lower TTFT (42\,ms vs.\
52\,ms). The two runtimes track each other up to concurrency~4
(96\,ms vs.\ 89\,ms). Beyond that, vLLM's TTFT inflates rapidly
(c=8: 215\,ms; c=16: 288\,ms; c=32: 759\,ms), while TensorRT-LLM's
PyTorch backend scales much more gradually (c=8: 142\,ms; c=16:
240\,ms; c=32: 319\,ms). At the highest concurrency tested, vLLM's
p50 TTFT is $2.4\times$ TensorRT-LLM's. The same trend holds in
the tails: at c=32, p95 TTFT is 1.30\,s for vLLM and 1.58\,s for
TensorRT-LLM (Figure~\ref{fig:ttft-concurrency}). This contradicts
the headline ``vLLM is always faster'' result of
Section~\ref{sec:seq}, and is the single most consequential
finding in the paper: \emph{the leadership flips with load}.

\paragraph{The flip is not explained by cache effectiveness.} The
expected mechanism for a flip---TensorRT-LLM's manual warning that
in-flight prefill blocks reuse---would predict a drop in
TensorRT-LLM's cache hit rate at high concurrency. We do not
observe that drop: the cache itself works equally well in both
runtimes at every concurrency we tested (Section~\ref{sec:hitrate}),
which rules out cache effectiveness as the explanation.

\paragraph{Queueing delay is a growing share of burst-mode TTFT.}
vLLM's counters separate scheduling wait
(\texttt{request\_queue\_time\_seconds}) from prefill compute, and
the per-cell deltas we already collect (Section~\ref{sec:driver})
localize its burst TTFT inflation: mean queueing delay rises from
0.0\,ms at c=1 to 11\,ms (c=8), 51\,ms (c=16), and 154\,ms
(c=32)---0.02\,\% to 20.3\,\% of p50 TTFT---while mean prefill
grows only $11\times$. Under fixed-rate arrival the same counter
stays below 0.1\,ms through 16\,QPS, consistent with vLLM's p95
tracking p50 within 2--3\,ms. Where we can measure directly,
scheduling delay rather than cache behavior varies with arrival
pattern. TensorRT-LLM exposed no comparable counters in the tested
version, so the cross-runtime \emph{flip} remains an inference
from cache-effect elimination plus the batching plateaus of
Figure~\ref{fig:cdf}; we present the scheduling account as the
best-supported hypothesis, not a fully instrumented causal claim
(Section~\ref{sec:limits}).

\subsection{Cache hit rate under burst arrival}
\label{sec:hitrate}

The two runtimes expose cache effectiveness through structurally
different native counters, not directly comparable without
normalization: vLLM reports block-granular hit/query counts at its
KV-cache block size (16 tokens by default), TensorRT-LLM a
per-request token count in the OpenAI-protocol response.
We normalize both to one quantity, the fraction of input tokens
served from cache, i.e.\ tokens that did not require prefill. For
vLLM this is \texttt{prefix\_cache\_hits\_total} over
\texttt{prefix\_cache\_queries\_total} (identical block units, so
the ratio is already a token fraction); for TensorRT-LLM,
cumulative \texttt{usage.prompt\_tokens\_details.cached\_tokens}
over cumulative \texttt{prompt\_tokens}. Both answer the same
question, and we compute each offline from raw per-cell logs with
one shared formula rather than trusting either runtime's
self-reported aggregate. They agree closely enough to be treated
as the same quantity: in Section~\ref{sec:pressure} they differ by
0.01 percentage points on the most aggressive reuse cell
(68.17\,\% vs.\ 68.18\,\%) and are identical (3.62\,\%) in the
capped rerun. We report aggregate agreement only; a per-request
comparison is left to the artifact.

The two metrics agree closely (vLLM 0.85 at every measured
concurrency; TensorRT-LLM 0.85 at c=1, dropping marginally to
0.83 at c=32). The total drop is under 3 percentage points over a
32$\times$ concurrency range. We conclude that, with four distinct
shared prefixes and a 4096-token shared prefix, neither runtime's
cache structure is the bottleneck under burst arrival in our
configuration. The first-request blocking effect described in the
TensorRT-LLM manual is real but, with as few as four prefixes,
amortizes within the first burst; subsequent bursts find the cache
warm.

\begin{figure}[t]
  \centering
  \includegraphics[width=0.82\columnwidth]{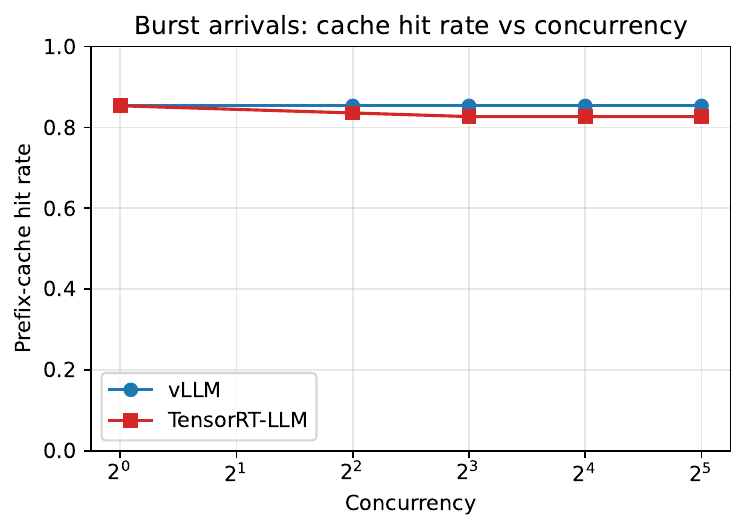}
  \caption{\textbf{Cache hit rate is essentially concurrency-insensitive.}
           Both runtimes hold $\approx$\,83--85\,\% hit rate across a
           32$\times$ concurrency range. The TensorRT-LLM manual's
           ``in-flight prefill blocks reuse'' caveat amortizes inside
           the first burst when num\_prefixes is small.
           Reporting paths: vLLM \texttt{/metrics} deltas;
           TensorRT-LLM \texttt{usage.prompt\_tokens\_details.cached\_tokens}.}
  \label{fig:hitrate}
\end{figure}

\subsection{TensorRT-LLM \texttt{tokens\_per\_block} ablation}
\label{sec:tpb}

The TensorRT-LLM manual describes \texttt{tokens\_per\_block} as a
tradeoff between reuse granularity and attention-kernel efficiency,
with default value~32. To check whether this tradeoff is exposed by
the PyTorch backend, we hold the workload constant (prefix length
4096, four prefixes, suffix length 512, output length 128, 100
requests per cell) and sweep \texttt{tokens\_per\_block} over
$\{32,64,128\}$ at three concurrencies $\{1,8,32\}$.

The result is essentially a null: across all nine cells, p50 TTFT
varies by under 7\,ms (well within run-to-run noise) and cache
hit rate is identical to four decimal places. At concurrency~1,
TTFT is 51.6\,ms, 51.7\,ms, and 51.9\,ms for tokens-per-block
32, 64, and 128 respectively; at concurrency~32, the same three
values give 288.9, 298.6, and 284.9\,ms. We conclude that, in the
PyTorch backend on H100 NVL with our workload shape, the
documented block-size tradeoff is not detectable above measurement
noise. Operators using the PyTorch backend can safely accept the
default and direct their tuning attention elsewhere.

\begin{figure}[t]
  \centering
  \includegraphics[width=0.82\columnwidth]{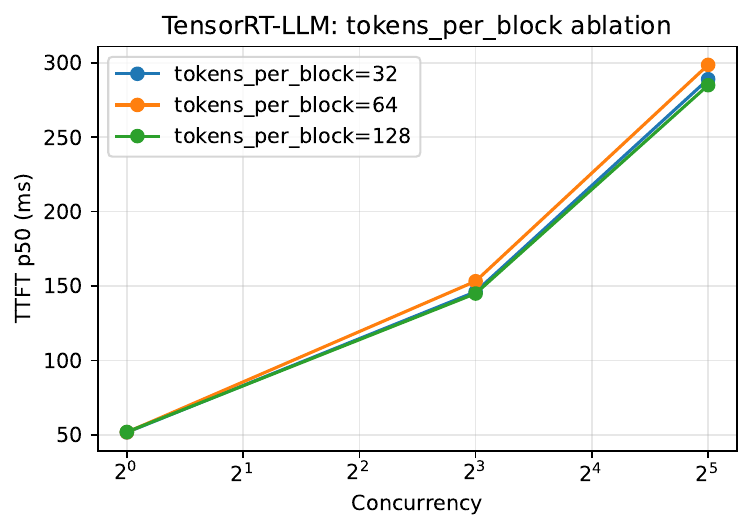}
  \caption{\textbf{Null result: \texttt{tokens\_per\_block} is not a useful
           tuning knob in the TRT-LLM PyTorch backend on H100.}
           The three settings (32, 64, 128) are indistinguishable to
           within $<$\,7\,ms at every concurrency. The
           reuse-granularity vs.\ kernel-efficiency tradeoff documented
           in the manual is not detectable above run-to-run noise.}
  \label{fig:tpb}
\end{figure}

\subsection{TensorRT-LLM engine-backend comparison}
\label{sec:enginebackend}

Every other result here uses TensorRT-LLM's PyTorch backend, the
documented default for TRT-LLM~1.x (Section~\ref{sec:method}).
TensorRT-LLM also ships a legacy engine backend that ahead-of-time
compiles the model. To test whether our findings are an artifact
of that path, we rerun the most aggressive reuse cell (pl=8192,
np=128, 384 requests) unchanged with \texttt{-{}-backend tensorrt}
(engine build 22.7\,s).

\begin{table}[t]
  \centering
  \caption{TensorRT-LLM PyTorch backend vs.\ engine backend,
           identical np=128 reuse cell (same cell as
           Table~\ref{tab:pressurecap}, uncapped).}
  \label{tab:enginebackend}
  \footnotesize
  \begin{tabular}{lcc}
    \toprule
    Backend & p50 TTFT & p95 TTFT \\
    \midrule
    PyTorch (default, used throughout) & 0.067\,s & 0.314\,s \\
    TensorRT engine                    & 0.069\,s & 0.661\,s \\
    \bottomrule
  \end{tabular}
\end{table}

Median TTFT is within 2\,ms between backends (0.067\,s vs.\
0.069\,s), consistent with comparable reuse behavior in this cell:
a full 8192-token prefill costs $\approx$\,0.3\,s
(Table~\ref{tab:reuse-speedup}), so a 69\,ms median is only
reachable with reuse active. Tail latency, by contrast, nearly
doubles on the engine backend (0.314\,s $\to$ 0.661\,s p95),
consistent with this paper's claim that cross-runtime differences
live in execution and scheduling above the cache rather than in
reuse effectiveness. This is a single-cell check, not a full re-validation: we did not
repeat the burst (Section~\ref{sec:burst}) or RAG
(Section~\ref{sec:rag}) sweeps on the engine backend, whose beta
status also means its \texttt{tokens\_per\_block} sensitivity may
differ from Section~\ref{sec:tpb}. Every other TensorRT-LLM result
here is scoped to the PyTorch backend and the tested 1.2.1
release; the benchmark methodology
(Sections~\ref{sec:method}--\ref{sec:sweeps}) is backend-agnostic,
and the \texttt{BACKEND} runner flag ships in the artifact.

\subsection{Realistic RAG-style workload}
\label{sec:rag}

\begin{figure}[t]
  \centering
  \includegraphics[width=0.82\columnwidth]{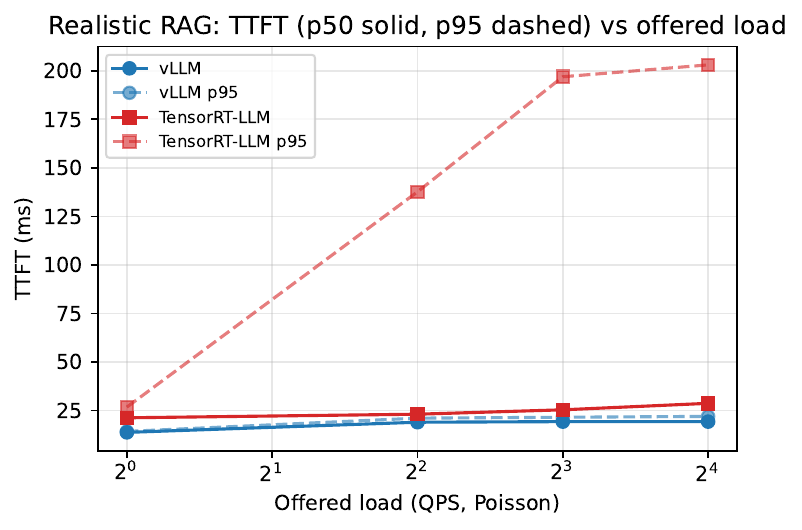}
  \caption{\textbf{Under steady fixed-rate load, the TTFT leadership
           flips back to vLLM by 9.2$\times$ on p95.} Solid lines are
           p50, dashed are p95. vLLM's p95 stays within 2--3\,ms of
           p50 across the swept QPS range; TensorRT-LLM's p95
           inflates by an order of magnitude (27\,ms $\to$ 203\,ms).
           The two runtimes are essentially equivalent on p50 but
           differ enormously on the tail.}
  \label{fig:realistic}
\end{figure}

Bursty arrival is a worst-case stressor; real serving traces are
steadier. We re-run the comparison using the template generator
(Section~\ref{sec:workloads}) with fixed-rate arrivals at
$\{1, 4, 8, 16\}$\,QPS and 200 requests per cell. Evenly spaced
arrivals are an idealisation of steady load: production traffic is
closer to Poisson, whose inter-arrival variance produces short-term
clustering that this pattern does not reproduce.
Figure~\ref{fig:realistic} plots p50 (solid) and p95 (dashed) TTFT
for both runtimes.

\begin{figure}[t]
  \centering
  \includegraphics[width=0.82\columnwidth]{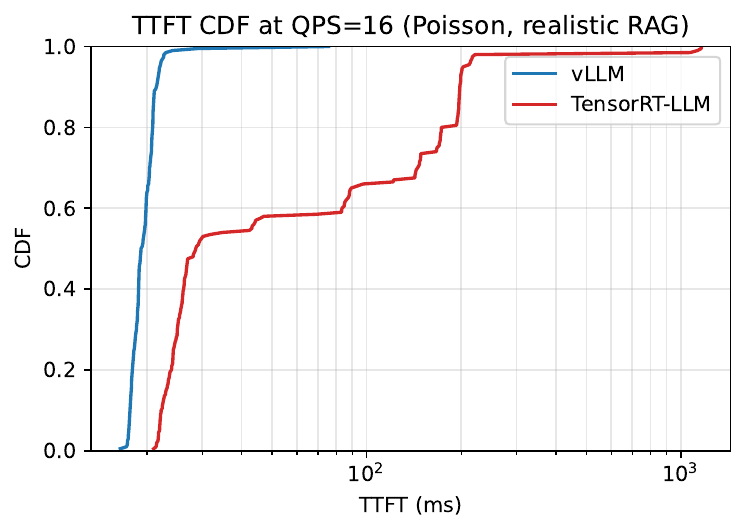}
  \caption{\textbf{TTFT CDF at QPS=16, realistic RAG workload.}
           vLLM is a tight near-vertical CDF around 20\,ms;
           TensorRT-LLM has two visible plateaus suggesting
           scheduler-induced batching at $\approx$\,200\,ms and
           $\approx$\,1\,s.}
  \label{fig:cdf}
\end{figure}

The story flips back. Under fixed-rate arrivals, vLLM dominates at
every offered load: p50 TTFT is 14, 19, 19, and 19\,ms at
QPS 1, 4, 8, 16 respectively, with p95 staying within 2-3\,ms of
p50 (14, 21, 22, 22\,ms). TensorRT-LLM's p50 climbs from 21\,ms
at QPS 1 to 29\,ms at QPS 16, but its \emph{p95} climbs much more
steeply: 27, 138, 197, and 203\,ms across the same range. At
QPS 16, TensorRT-LLM's p95 TTFT is $7.1\times$ its own p50 and
$9.2\times$ vLLM's p95. The mechanism behind both ratios is visible
in Figure~\ref{fig:cdf}: TensorRT-LLM's TTFT distribution is bimodal,
with a fast mode near its p50 and a plateau near 200\,ms that about a
tenth of requests land on, while vLLM's is unimodal and tight. Because
the p95 sits near the knee between TensorRT-LLM's two modes, its exact
value is sensitive to how many requests land on the plateau: an
independent re-execution during artifact evaluation, on different
H100 hardware, measured the same ordering but a smaller ratio. The
bimodality and the ordering are the robust findings; the multiplier is
specific to this hardware and configuration.

The cache itself continues to perform well in both runtimes (93\%
hit rate or above in all four realistic cells). The performance
gap is again upstream of the cache; in the realistic regime it is
TensorRT-LLM's tail-latency behavior under inter-arrival jitter
that dominates the user-visible serving cost.

Combined with the burst sweep, two arrival regimes give two
different rankings. The burst sweep had TensorRT-LLM ahead on p50
TTFT at concurrency~32; the realistic sweep has vLLM ahead by
$9\times$ on p95 at QPS 16. Operators tuning a system around the
arrival pattern they actually see in production cannot extrapolate
from either microbenchmark alone.

Figure~\ref{fig:cdf} shows the full TTFT distribution for the
QPS=16 cell of each runtime. vLLM's CDF is a tight near-vertical
line at $\approx$\,20\,ms; TensorRT-LLM's CDF has two visible
plateaus---one at $\approx$\,200\,ms (40\,\% of requests are above
this) and a second at $\approx$\,1\,s. The structure of the
TensorRT-LLM distribution suggests scheduler-induced batching
plateaus, not random tail noise.

\subsection{Output length is decoupled from prefix-reuse behavior}
\label{sec:olen}

The abstract motivated a regime in which long generations erode
prefix-reuse gains via increased KV-cache pressure. We test this
directly by holding the workload constant (4096-token prefix,
512-token suffix, single shared prefix, sequential arrival) and
sweeping output length over $\{32,128,512\}$ tokens.

The effect is null in our setting. p50 TTFT moves by at most
1.2\,ms across the three output lengths (vLLM: 39.7--40.9\,ms;
TensorRT-LLM: 51.0--51.3\,ms), under 3\,\% of TTFT against the
5--6.5$\times$ effect of reuse itself, and the
prefix-cache hit rate is identical to three decimal places at
0.859 for TensorRT-LLM and equivalent on vLLM. Throughput
\emph{does} scale with output length because TTFT overhead
amortizes over more decode tokens, but this is a property of the
TTFT-to-decode ratio, not of prefix reuse.

The KV-cache pressure regime the abstract mentioned thus requires
a different operating point than long generations alone: a large
number of \emph{distinct} prefixes that collectively exceed HBM
budget. Section~\ref{sec:pressure} addresses this directly.

\subsection{Cache pressure regime: can we break the cache on H100~NVL?}
\label{sec:pressure}

The findings so far come from modest cache-pressure operating
points (1--4 distinct prefixes). Sweeping the number of distinct
8192-token prefixes over $\{16, 64, 128\}$ with Qwen2-7B produces
no eviction in any cell, but this is a property of the operating
point, not of prefix caching: our GPU is a 94\,GB
\textbf{H100~NVL}, and at 7\,B in BF16 ($\approx$\,15\,GB of
weights) the KV pool exceeds even our largest working set
(1.05\,M tokens vs.\ $\geq$\,1.25\,M). The cache is never full, so
``no evictions'' is that arithmetic's expected outcome rather than
an empirical finding. The experiments below isolate the ratio of
prefix working set to KV capacity as the governing variable,
holding traces, prefix/suffix/output length, arrival pattern
(burst), and concurrency (1) identical to the original sweep;
Qwen2 and Qwen2.5 share a tokenizer, so token counts are exactly
comparable.

\paragraph{Model-scale pressure: Qwen2.5-32B-Instruct.} A larger
model shrinks the KV pool directly, since more HBM goes to weights.
We repeat the pressure sweep unchanged except for the served model,
substituting Qwen2.5-32B-Instruct ($\approx$\,65\,GB of BF16
weights). Runtime-reported KV capacity falls to 79{,}280 tokens
(4{,}955 vLLM blocks at 16 tokens/block; 19.4\,GB), roughly
$16\times$ smaller than the 7B pool, and we add a below-capacity
cell (np=4) as a same-model baseline. Working set is
number-of-prefixes~$\times$~8192; Table~\ref{tab:pressure32b}
reports hit rate and TTFT as the ratio of working set to measured
capacity grows from 0.4$\times$ to 13.2$\times$.

\begin{table}[t]
  \centering
  \caption{Cache pressure under model-scale KV shrinkage
           (Qwen2.5-32B-Instruct, 79{,}280-token / 19.4\,GB KV pool,
           94\,GB H100~NVL). Ratio is working set
           (np~$\times$~8192 tokens) over measured KV capacity.}
  \label{tab:pressure32b}
  \scriptsize
  \begin{tabular}{cccccccc}
    \toprule
    & & & \multicolumn{2}{c}{Hit rate} & \multicolumn{2}{c}{p50 TTFT} \\
    \cmidrule(lr){4-5}\cmidrule(lr){6-7}
    np & Ratio & Ceil. & vLLM & TRT & vLLM & TRT \\
    \midrule
    4   & 0.4$\times$  & 91.7\,\% & 88.9\,\% & 88.9\,\% & 0.091\,s & 0.083\,s \\
    16  & 1.7$\times$  & 66.7\,\% & 46.6\,\% & 52.5\,\% & 1.075\,s & 0.109\,s$^\dagger$ \\
    64  & 6.6$\times$  & 66.7\,\% & 12.6\,\% & 13.8\,\% & 1.420\,s & 1.394\,s \\
    128 & 13.2$\times$ & 66.7\,\% & 5.6\,\%  & 6.6\,\%  & 1.421\,s & 1.398\,s \\
    \bottomrule
  \end{tabular}
  \\[2pt]
  \raggedright\footnotesize ``Ceil.'' is the analytical no-eviction
  ceiling $1-\mathrm{np}/N$ for a cell of $N$ requests drawing
  uniformly from np prefixes: the hit rate a perfect, never-evicting
  cache would achieve, since each prefix must miss once. The np=4
  baseline uses 48 requests (12 revisits per prefix) and the
  remaining cells 3 revisits per prefix, so ceilings differ and raw
  hit rates are \emph{not} directly comparable across rows; the
  eviction test is each cell against its own ceiling.
  $^\dagger$At $\approx$\,50\,\% hit rate the median sits on the
  hit/miss boundary, so this p50 is unstable and we draw no
  cross-runtime conclusion from it.
\end{table}

Three findings follow. First, the below-capacity cell sits at its
no-eviction ceiling (88.9\,\% against 91.7\,\%) while every
over-subscribed cell falls far below its own ceiling: 46.6\,\%
against 66.7\,\% at 1.7$\times$, and 5.6\,\% against the same
66.7\,\% at 13.2$\times$. Comparing each cell to its own ceiling
rather than to the other rows isolates eviction from the
revisit-count difference between them: this \emph{is} the eviction
regime the paper motivated but had not previously reached. Neither
runtime exposes a block-eviction counter, so we identify the regime
by this shortfall rather than by counting evictions directly: a
never-evicting cache cannot fall below its ceiling, because every
revisit would hit. Second,
the shortfall below ceiling deepens progressively with
over-subscription rather than at a single threshold, and p50
approaches p95 at the two deepest cells, meaning essentially every
request pays a full prefill because its prefix was evicted before
its next revisit. Third, and central to this paper's argument, the
caching-\emph{parity} finding of Section~\ref{sec:seq} survives
under pressure: the two runtimes' hit rates track within 1--6
percentage points at every ratio, as they did below capacity.
Model scale changes \emph{whether} the cache is under pressure,
not \emph{which runtime} caches better.

\paragraph{Harness validation: capacity-capped Qwen2-7B.} The 32B
experiment changes the model, and so per-request compute cost. To
isolate KV \emph{capacity}, we return to Qwen2-7B and cap the pool
at 56{,}320 tokens, below the measured 32B pool, via each
runtime's own control (\texttt{num\_gpu\_blocks\_override=3520};
\texttt{KvCacheConfig.max\_tokens=56320}), then rerun the np=128
cell unchanged (Table~\ref{tab:pressurecap}).

\begin{table}[t]
  \centering
  \caption{Capacity-capped Qwen2-7B, np=128, pl=8192 (same cell as
           the original submission). Capping KV capacity to
           56{,}320 tokens on both runtimes reproduces the eviction
           regime independent of model choice.}
  \label{tab:pressurecap}
  \scriptsize
  \begin{tabular}{llccc}
    \toprule
    Runtime & KV cap & Hit & p50 & p95 \\
    \midrule
    vLLM    & uncapped ($\geq$1.25M tok) & 68.2\,\% & 0.052\,s & 0.329\,s \\
    vLLM    & 56.3k tok                  & 3.6\,\%  & 0.318\,s & 0.328\,s \\
    \midrule
    TRT-LLM & uncapped ($\geq$1.25M tok) & 68.2\,\% & 0.067\,s & 0.314\,s \\
    TRT-LLM & 56.3k tok                  & 3.6\,\%  & 0.303\,s & 0.312\,s \\
    \bottomrule
  \end{tabular}
\end{table}

Capping capacity collapses the hit rate from 68.2\,\% to 3.6\,\% on
\emph{both} runtimes, identical to two decimal places, with a
4.6--6.1$\times$ TTFT-p50 regression. The harness therefore reaches
the eviction regime whenever the prefix working set exceeds
configured KV capacity, independent of the model served; the
earlier null result reflects the 7B/94\,GB operating point, not a
limitation of the methodology.

\paragraph{Takeaway.} Regime onset is driven primarily by the ratio
of prefix working set to available KV capacity rather than by
absolute model size: a capped 7B pool and a natural 32B pool reach
the same eviction behavior at comparable ratios. Larger models,
longer contexts, higher concurrency, and multi-tenancy all shrink
that denominator and should push deployments into this regime
sooner than our 7B baseline suggests. 70B-class and multi-GPU
serving, where KV capacity and cross-device scheduling interact in
ways a single-H100 study cannot characterize, remain future work.

\subsection{Cross-model generalization: Mistral-7B-Instruct-v0.3}
\label{sec:crossmodel}

This section and Section~\ref{sec:pressure} test two distinct
robustness axes that should not be conflated. Here we hold
parameter count fixed at $\approx$\,7\,B and vary
\emph{architecture} (attention layout, layer count, tokenizer),
repeating the burst sweep on Mistral-7B-Instruct-v0.3 (GQA with
eight KV heads, vs.\ Qwen2's four) with an identical
configuration: 4096-token shared prefix, four distinct prefixes,
100 requests per cell at concurrencies $\{1, 8, 32\}$.
Section~\ref{sec:pressure} instead holds architecture fixed and
varies \emph{scale} (7\,B vs.\ 32\,B). Neither check substitutes
for the other.

The findings replicate cleanly (Table~\ref{tab:crossmodel}).
At concurrency~1, both runtimes are within 3\,ms of the Qwen2
numbers in the same direction (vLLM lower than TensorRT-LLM). At
concurrency~32, the TTFT-leadership flip survives: vLLM has p50
TTFT 553.8\,ms vs TensorRT-LLM's 265.0\,ms ($2.1\times$ vLLM/TRT
ratio, compared to $2.4\times$ on Qwen2). The cache hit rate at
c=1 is 85\,\%, identical to Qwen2 to two decimal places. We
conclude that the central findings of this paper are properties of
the runtimes interacting with H100~NVL, not properties of the
Qwen2 architecture.

\begin{table}[t]
  \centering
  \caption{Cross-model verification of the TTFT-leadership flip
           (p50 TTFT, ms). The flip survives the architecture
           change.}
  \label{tab:crossmodel}
  \footnotesize
  \begin{tabular}{lcccc}
    \toprule
    Model & Runtime & c=1 & c=8 & c=32 \\
    \midrule
    Qwen2-7B    & vLLM         & 42.2 & 214.7 & 758.6 \\
    Qwen2-7B    & TRT-LLM      & 51.7 & 142.1 & 319.1 \\
    \midrule
    Mistral-7B  & vLLM         & 40.0 & 195.6 & 553.8 \\
    Mistral-7B  & TRT-LLM      & 49.8 & 161.2 & 265.0 \\
    \midrule
    \multicolumn{2}{l}{vLLM / TRT ratio (Qwen2)}     & 0.82 & 1.51 & \textbf{2.38} \\
    \multicolumn{2}{l}{vLLM / TRT ratio (Mistral)}   & 0.80 & 1.21 & \textbf{2.09} \\
    \bottomrule
  \end{tabular}
\end{table}

\subsection{Energy savings from prefix reuse}
\label{sec:energy}

Beyond latency and throughput, prefix reuse should also reduce
\emph{energy} consumed per request, because the avoided prefill
compute would otherwise drive the GPU at higher power. To
quantify this, we sample \texttt{nvidia-smi --query-gpu=power.draw}
at 4\,Hz during each cell of a focused energy sweep
(40 sequential requests at 8192-token shared prefix, with reuse
on vs.\ off, both runtimes).

Table~\ref{tab:energy} reports the result. Reuse reduces average
power draw by 9--10\,\% (because expensive prefill kernels are
replaced by inexpensive cache lookups) \emph{and} shortens wall
time by 5--6$\times$ (Table~\ref{tab:reuse-speedup}). The net
energy-per-request savings are 24\,\% (TensorRT-LLM) and 31\,\%
(vLLM); on a per-token basis the same numbers apply because output
token count is fixed.

\begin{table}[t]
  \centering
  \caption{Average GPU power and energy per request at pl=8192,
           c=1, 40 requests. Reuse delivers both lower power draw
           and shorter wall time, compounding to 24--31\,\%
           energy savings.}
  \label{tab:energy}
  \footnotesize
  \begin{tabular}{lccc}
    \toprule
    Configuration       & Avg power & Wall time & Energy/req \\
    \midrule
    vLLM, reuse on      & 259.2\,W & $\approx$\,52\,s & 337\,J \\
    vLLM, reuse off     & 285.9\,W & $\approx$\,62\,s & 443\,J \\
    \rowcolor[gray]{0.95}
    vLLM savings        & 9.3\,\%  & 16\,\%           & \textbf{24--31\,\%} \\
    \midrule
    TRT-LLM, reuse on   & 238.3\,W & $\approx$\,67\,s & 399\,J \\
    TRT-LLM, reuse off  & 261.2\,W & $\approx$\,76\,s & 496\,J \\
    \rowcolor[gray]{0.95}
    TRT-LLM savings     & 8.8\,\%  & 12\,\%           & \textbf{20--24\,\%} \\
    \bottomrule
  \end{tabular}
\end{table}

For a deployment serving 1\,M long-prefix requests/day, this
24--31\,\% energy savings corresponds to several tens of kWh/day at
H100-class power draw---enough to make prefix reuse a sustainability
optimization in addition to a latency one.

\subsection{Run-to-run stability and the (prefix, concurrency) operating map}
\label{sec:matrix}

To verify that the headline numbers are not driven by noise we
re-run the pl=8192 concurrency=1 cell three times for each runtime.
The measured p50 TTFTs are 54.0, 53.8, 54.0\,ms (vLLM,
$\sigma$=0.13\,ms) and 58.3, 58.0, 57.9\,ms (TensorRT-LLM,
$\sigma$=0.21\,ms), variation under 0.7\,\% within each runtime
and a cross-runtime gap that is more than 20\,$\sigma$.

\begin{figure}[t]
  \centering
  \includegraphics[width=0.82\columnwidth]{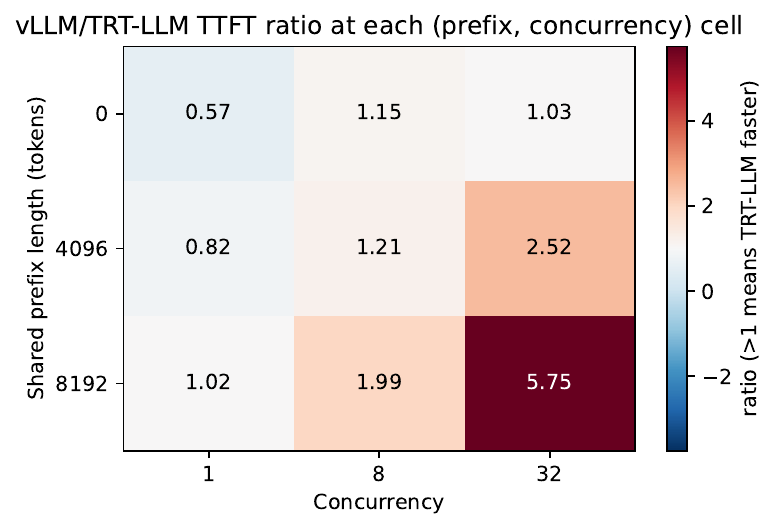}
  \caption{\textbf{The TTFT-leadership crossover is a smooth surface
           on the (prefix length, concurrency) plane.}
           Cells show the vLLM/TRT-LLM p50 TTFT ratio; values below
           1 mean vLLM is faster and above 1 mean TensorRT-LLM is
           faster, with darker cells indicating a larger ratio. The
           leadership flips along the diagonal, and the TensorRT-LLM
           advantage at high concurrency grows with prefix length,
           peaking at 5.75$\times$ at the (pl=8192, c=32) corner.}
  \label{fig:matrix}
\end{figure}

To map the full operating envelope we sweep prefix length
$\in \{0, 4096, 8192\}$ tokens against concurrency
$\in \{1, 8, 32\}$ for both runtimes (80 requests per cell, 4
distinct prefixes). Figure~\ref{fig:matrix} reports the vLLM/TRT-LLM
p50 TTFT ratio at each cell.

The map has a clean diagonal: along the top-left edge (short
prefix, low concurrency) vLLM leads (ratio $<$\,1, blue), while
along the bottom-right edge (long prefix, high concurrency)
TensorRT-LLM dominates by a factor that grows with both axes. At
the extreme corner (pl=8192, c=32), the ratio is 5.75$\times$---a
larger crossover than the 2.4$\times$ gap reported in
Section~\ref{sec:burst} at pl=4096. The TTFT-leadership crossover
is not a single threshold but a smooth surface; long prefixes amplify
the effect of high concurrency.

\subsection{Throughput envelope}
\label{sec:tput}

The throughput envelope tells the opposite story to TTFT. At
concurrency~1 the two runtimes are within 30\,\% of each other
(vLLM 129\,t/s, TensorRT-LLM 92\,t/s), but as concurrency grows,
vLLM's output token rate scales steeply while TensorRT-LLM's
PyTorch backend saturates. At c=32, vLLM produces 1444\,t/s vs.\
TensorRT-LLM's 523\,t/s---a $2.8\times$ throughput advantage at
the same concurrency point where TensorRT-LLM had the
$2.4\times$ TTFT advantage. Operators face a real tradeoff: the
runtime that minimizes p50 first-token latency under burst load
is not the same runtime that maximizes raw token throughput.

\begin{figure}[t]
  \centering
  \includegraphics[width=0.82\columnwidth]{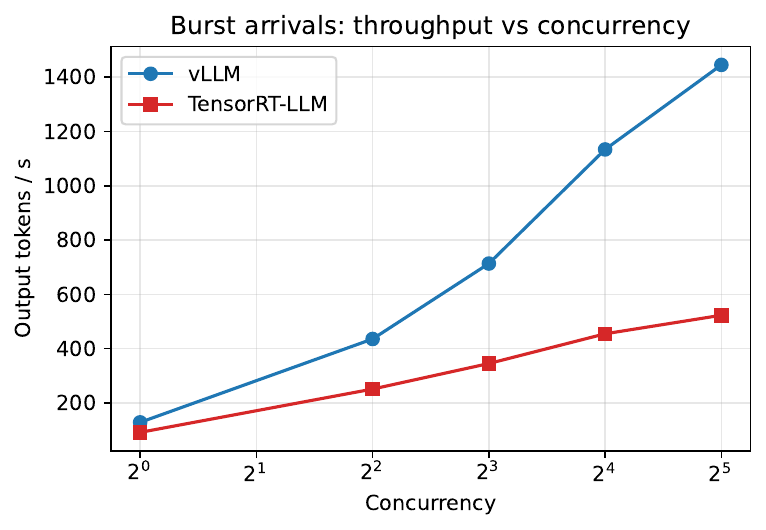}
  \caption{\textbf{Throughput tells the opposite story to TTFT: vLLM
           dominates at high concurrency by 2.8$\times$.}
           Output tokens/s vs.\ concurrency, burst arrival, 4096-token
           shared prefix. At c=32, vLLM delivers 1444\,t/s vs.\
           TensorRT-LLM's 523\,t/s---the same concurrency point where
           TensorRT-LLM had the 2.4$\times$ TTFT advantage in
           Figure~\ref{fig:ttft-concurrency}.}
  \label{fig:throughput}
\end{figure}

\section{Discussion and Tuning Guidance}
\label{sec:discussion}

We synthesize the findings into three operator-facing observations.

\paragraph{Observation 1: cache effectiveness is solved
\emph{below} the pressure regime, and the two runtimes remain tied
even inside it.} Both runtimes cache 92\% of the input at an
8192-token shared prefix (Section~\ref{sec:seq}), and the hit rate
stays within 3 percentage points across a 32$\times$ concurrency
range (Section~\ref{sec:hitrate}). At the prefix-pool sizes of
Sections~\ref{sec:seq}--\ref{sec:rag}, the 94\,GB HBM headroom
keeps the working set inside KV capacity and vLLM's
\texttt{kv\_cache\_usage\_perc} under 1\,\%. That headroom is
finite: past capacity (Section~\ref{sec:pressure}), hit rate
degrades sharply for both runtimes together and parity still
holds. Reuse is free only below a deployment's
working-set/capacity ratio; above it, both runtimes need the same
capacity planning, not a runtime-specific fix.

\paragraph{Observation 2: Arrival pattern, not cache configuration,
selects the better runtime.} The most consequential cross-runtime
difference is not in the cache but in how each scheduler handles
burst vs.\ fixed-rate traffic. Under
burst arrival at c=32, TensorRT-LLM has $2.4\times$ lower p50 TTFT
(Section~\ref{sec:burst}); under fixed-rate arrival at 16\,QPS, vLLM
has $9.2\times$ lower p95 TTFT (Section~\ref{sec:rag}). Neither
result generalizes to the other regime. Operators should therefore
benchmark with the arrival pattern they actually see in
production, not the synthetic burst harness shipped with either
runtime.

\paragraph{Observation 3: \texttt{tokens\_per\_block} in the
TensorRT-LLM PyTorch backend is not a useful knob in this regime.}
The documented reuse-granularity vs.\ kernel-efficiency tradeoff is
not detectable above noise across $\{32,64,128\}$ at any
concurrency we tested (Section~\ref{sec:tpb}). Tuning effort is
better spent on \texttt{max\_batch\_size} and arrival-pattern
shaping at the load balancer.

Table~\ref{tab:tuning} summarizes the per-workload recommendation.

\begin{table}[t]
  \centering
  \caption{Recommended runtime configuration by workload shape on
           a single H100~NVL, from \textsc{PrefixBench-H100}. All
           settings assume Qwen2-7B-Instruct in BF16; reuse is
           enabled by default in both runtimes.}
  \label{tab:tuning}
  \footnotesize
  \begin{tabular}{p{0.27\linewidth}p{0.28\linewidth}p{0.31\linewidth}}
    \toprule
    Workload pattern & Preferred runtime & Notes \\
    \midrule
    Long shared system prompt, sequential / interactive
      & vLLM (lower TTFT and ITL)
      & Both runtimes cache identically; vLLM has $\approx$\,8\,ms
        less per-request overhead. \\
    Bursty short prompts, p50 TTFT-critical
      & TensorRT-LLM at c $\geq$ 16
      & vLLM's TTFT inflates 2$\times$--3$\times$ faster under burst;
        TensorRT-LLM's scheduler scales better. \\
    Steady fixed-rate load (RAG, agent loops)
      & vLLM (much lower p95)
      & TensorRT-LLM's PyTorch backend exhibits a 9$\times$ p95
        inflation between 1\,QPS and 16\,QPS that vLLM does not. \\
    Throughput-maximizing batch (offline)
      & vLLM
      & 1444\,t/s vs.\ 523\,t/s at c=32 in our setup. \\
    Block-size tuning (TRT-LLM)
      & default (32)
      & No detectable difference at $\{32,64,128\}$. \\
    \bottomrule
  \end{tabular}
\end{table}

\subsection{Limitations}
\label{sec:limits}

The results are limited to the configurations we tested.
(0)~Our steady-load cells use \emph{fixed-rate} arrivals evenly
spaced at $1/\mathrm{QPS}$ (the zero-variance idealisation of
steady traffic) rather than Poisson arrivals. Poisson traffic adds
short-term clustering, which we expect would move steady-load tail
latencies toward the burst-arrival results, so our
burst-versus-steady contrast spans a wider arrival-variance range
than production traffic typically exhibits.
(i)~Two model scales (7\,B and 32\,B,
Section~\ref{sec:pressure}), both BF16; we do not evaluate
70\,B-class models, multi-GPU serving, or FP8 KV cache, all of
which shrink per-device KV capacity and could shift the pressure
regime sooner than our results suggest.
(ii)~All measurements use a single 94\,GB H100~NVL. The pressure
regime is reachable here once the working-set-to-capacity ratio
passes parity, but we have not measured whether the ratio at which
degradation begins differs on smaller-HBM Hopper variants (e.g.\
H100~PCIe/SXM at 80\,GB).
(iii)~TensorRT-LLM's PyTorch backend is the primary evaluation
target throughout, and the \texttt{tokens\_per\_block} ablation
(Section~\ref{sec:tpb}) is PyTorch-only; the single-cell engine-backend
comparison (Section~\ref{sec:enginebackend}) is consistent with
reuse parity but is not a full re-validation.
(iv)~The OpenAI HTTP transport adds $\approx$\,2--3\,ms of fixed
overhead shared by both runtimes; it does not affect the
cross-runtime comparison but does affect absolute TTFT.
(v)~The scheduling account of Section~\ref{sec:burst} is directly
measured for vLLM (queueing vs.\ prefill decomposition) but rests
on hit-rate elimination and CDF-shape evidence for the
cross-runtime flip, since TensorRT-LLM exposed no comparable
scheduler counters in the tested version; we present it as the
best-supported hypothesis, not a fully instrumented causal claim.
(vi)~Nsight traces beyond the case-study figure are deferred to
follow-up work.

\section{Related Work}
\label{sec:related}

\textsc{PrefixBench-H100} sits at the intersection of three lines of
prior work: KV-cache memory management, phase-aware serving system
design, and accelerator-level workload characterization.

\paragraph{KV-cache memory management.} PagedAttention introduced
fixed-size blocks and dynamic block tables to eliminate KV-cache
fragmentation and enable block-level sharing across
requests~\cite{kwon2023pagedattention}. vAttention argues for
virtual-memory-style management without paging, achieving similar
sharing without explicit block tables~\cite{prabhu2024vattention}.
SGLang's RadixAttention organizes cached prefixes in a radix tree
for fine-grained reuse across branching request structures rather
than a fixed block grid~\cite{zheng2024sglang}, and ChunkAttention
restructures the attention kernel around a prefix-aware KV
layout~\cite{ye2024chunkattention}. LMCache extends reuse beyond a
single GPU's HBM with a multi-tier cache
layer~\cite{cheng2025lmcache}, MemServe pools KV state across a
disaggregated deployment~\cite{hu2024memserve}, CacheGen
compresses and streams KV state to cut transfer
cost~\cite{liu2024cachegen}, and CacheBlend fuses cached chunks
for RAG without recomputing cross-chunk
attention~\cite{yao2025cacheblend}. All
focus on \emph{mechanism design}: new data structures, kernels, or
placement strategies for reuse. Our work is complementary: we
measure how the mechanisms vLLM and TensorRT-LLM already ship
perform on H100-class hardware, the empirical question these
papers motivate but leave unanswered for shipped, black-box
runtimes.

\paragraph{Phase-aware serving.} Orca introduced iteration-level
scheduling with selective batching, the continuous-batching
foundation that both runtimes we evaluate build
on~\cite{yu2022orca}. Sarathi-Serve tames the throughput-latency
tradeoff via chunked prefill that overlaps with ongoing decode
steps~\cite{agrawal2024sarathi}. DeepSpeed-FastGen combines
dynamic SplitFuse (a chunked-prefill variant) with a
continuous-batching engine to raise effective
throughput~\cite{deepspeedfastgen2024}.
DistServe~\cite{zhong2024distserve} and
Splitwise~\cite{patel2024splitwise} disaggregate prefill and decode
onto separate GPU pools, exploiting the same prefill/decode asymmetry.
Disaggregation is out of scope here, since we restrict ourselves to
a single H100, but the asymmetry these systems exploit is also what
makes prefix reuse so attractive: it converts a portion of an
expensive prefill into a cache lookup. This literature also frames
our scheduler finding (Section~\ref{sec:burst}): both runtimes
implement iteration-level continuous batching in the
Orca/Sarathi-Serve tradition, and our results indicate the
batch-admission choices layered on that foundation, not cache
effectiveness, determine which wins under a given arrival
pattern.

\paragraph{Throughput and quantization-aware serving.}
QServe~\cite{lin2024qserve} and NanoFlow~\cite{zhu2024nanoflow}
both push single-GPU throughput limits via quantization and
fine-grained intra-device scheduling. Quantization (FP8 / INT8 KV
cache) is itself a knob that interacts with prefix reuse on H100;
this paper restricts itself to BF16 and leaves the precision
interaction to follow-up work.

\paragraph{Workload characterization.}
The closest stylistic precedent in the IISWC literature is
LLMServingSim~\cite{cho2024llmservingsim}, a HW/SW co-simulation
infrastructure for LLM inference serving, and the distributed-LLM
workload analyses that have appeared in recent IISWC programs.
\textsc{PrefixBench-H100} differs from those works in three ways:
(i)~it targets a single H100 NVL with real measurements rather than
simulation, (ii)~it studies two production runtimes head-to-head
under matched workloads, and (iii)~it focuses specifically on prefix
reuse rather than the full serving design space.

\section{Conclusion}
\label{sec:conclusion}

\textsc{PrefixBench-H100} characterizes prefix reuse on a single
NVIDIA H100 NVL across two production LLM serving runtimes. Driving
vLLM and TensorRT-LLM through identical workloads via an
OpenAI-compatible load generator separates \emph{intrinsic}
prefix-reuse benefits from runtime-specific implementation effects.
The resulting operating map bounds the regime in which prefix reuse
is the single most effective TTFT optimization on H100: reuse
delivers 5--6.5$\times$ lower TTFT while the prefix working set fits
in KV capacity, and degrades to a 5.6\,\% hit rate and
15.7$\times$ TTFT regression at 13.2$\times$ over-subscription. In
every regime, including under eviction, the runtimes' cache
effectiveness stays within a couple of points, so the differences
operators see come from the scheduling layer above the cache. The
released artifact is intended to be extended to further model families,
runtimes, and Hopper-class accelerators, building a shared empirical
baseline for phase-aware LLM serving research.

\section*{Acknowledgement of Generative AI Use}
\label{sec:ai-ack}

As required by the IISWC~2026 author guidelines: we used
Anthropic's Claude as a coding and writing assistant to scaffold
the benchmark code in \texttt{prefixbench/}, drive the H100
sweeps, generate figures, and draft initial prose. All
experimental designs, workload-axis choices, interpretations, and
scientific claims were directed and verified by the authors. Every
data point comes from running the released code on the hardware of
Section~\ref{sec:method}; no measurement was synthesized or
fabricated. The tool \emph{assisted}, and did not replace, the
authors' judgement.

\bibliographystyle{IEEEtranS}
\bibliography{refs}

@inproceedings{kwon2023pagedattention,
  title     = {Efficient Memory Management for Large Language Model Serving with {PagedAttention}},
  author    = {Kwon, Woosuk and Li, Zhuohan and Zhuang, Siyuan and Sheng, Ying and Zheng, Lianmin and Yu, Cody Hao and Gonzalez, Joseph E. and Zhang, Hao and Stoica, Ion},
  booktitle = {Proceedings of the 29th ACM Symposium on Operating Systems Principles (SOSP '23)},
  year      = {2023}
}

@inproceedings{prabhu2024vattention,
  title     = {{vAttention}: Dynamic Memory Management for Serving {LLMs} without {PagedAttention}},
  author    = {Prabhu, Ramya and Nayak, Ajay and Mohan, Jayashree and Ramjee, Ramachandran and Panwar, Ashish},
  booktitle = {Proceedings of the 30th ACM International Conference on Architectural Support for Programming Languages and Operating Systems (ASPLOS '25)},
  year      = {2025}
}

@inproceedings{agrawal2024sarathi,
  title     = {Taming Throughput-Latency Tradeoff in {LLM} Inference with {Sarathi-Serve}},
  author    = {Agrawal, Amey and Kedia, Nitin and Panwar, Ashish and Mohan, Jayashree and Kwatra, Nipun and Gulavani, Bhargav S. and Tumanov, Alexey and Ramjee, Ramachandran},
  booktitle = {18th USENIX Symposium on Operating Systems Design and Implementation (OSDI '24)},
  year      = {2024}
}

@inproceedings{zhong2024distserve,
  title     = {{DistServe}: Disaggregating Prefill and Decoding for Goodput-optimized Large Language Model Serving},
  author    = {Zhong, Yinmin and Liu, Shengyu and Chen, Junda and Hu, Jianbo and Zhu, Yibo and Liu, Xuanzhe and Jin, Xin and Zhang, Hao},
  booktitle = {18th USENIX Symposium on Operating Systems Design and Implementation (OSDI '24)},
  year      = {2024}
}

@inproceedings{patel2024splitwise,
  title     = {Splitwise: Efficient Generative {LLM} Inference Using Phase Splitting},
  author    = {Patel, Pratyush and Choukse, Esha and Zhang, Chaojie and Shah, Aashaka and Goiri, {\'I}{\~n}igo and Maleki, Saeed and Bianchini, Ricardo},
  booktitle = {Proceedings of the 51st Annual International Symposium on Computer Architecture (ISCA '24)},
  year      = {2024}
}

@inproceedings{lin2024qserve,
  title     = {{QServe}: {W4A8KV4} Quantization and System Co-design for Efficient {LLM} Serving},
  author    = {Lin, Yujun and Tang, Haotian and Yang, Shang and Zhang, Zhekai and Xiao, Guangxuan and Gan, Chuang and Han, Song},
  booktitle = {Proceedings of Machine Learning and Systems (MLSys '25)},
  year      = {2025}
}

@inproceedings{zhu2024nanoflow,
  title     = {{NanoFlow}: Towards Optimal Large Language Model Serving Throughput},
  author    = {Zhu, Kan and Gao, Yufei and Zhao, Yilong and Zhao, Liangyu and Zuo, Gefei and Gu, Yile and Xie, Dedong and Tang, Tian and Xu, Qinyu and Ye, Zihao and Kamahori, Keisuke and Lin, Chien-Yu and Wang, Ziren and Wang, Stephanie and Krishnamurthy, Arvind and Kasikci, Baris},
  booktitle = {19th USENIX Symposium on Operating Systems Design and Implementation (OSDI '25)},
  year      = {2025}
}

@misc{nvidia_h100_datasheet,
  title  = {{NVIDIA H100 Tensor Core GPU Datasheet}},
  author = {{NVIDIA}},
  year   = {2023},
  note   = {Online; accessed 2026-05-10}
}

@misc{trtllm_docs,
  title  = {{TensorRT-LLM Documentation}},
  author = {{NVIDIA}},
  note   = {\url{https://nvidia.github.io/TensorRT-LLM/}; accessed 2026-05-10}
}

@inproceedings{cho2024llmservingsim,
  title     = {{LLMServingSim}: A {HW/SW} Co-Simulation Infrastructure for {LLM} Inference Serving at Scale},
  author    = {Cho, Jaehong and Kim, Minsu and Choi, Hyunmin and Heo, Guseul and Park, Jongse},
  booktitle = {Proceedings of the IEEE International Symposium on Workload Characterization (IISWC)},
  year      = {2024}
}

@inproceedings{zheng2024sglang,
  title     = {{SGLang}: Efficient Execution of Structured Language Model Programs},
  author    = {Zheng, Lianmin and Yin, Liangsheng and Xie, Zhiqiang and Sun, Chuyue and Huang, Jeff and Yu, Cody Hao and Cao, Shiyi and Kozyrakis, Christos and Stoica, Ion and Gonzalez, Joseph E. and Barrett, Clark and Sheng, Ying},
  booktitle = {Advances in Neural Information Processing Systems (NeurIPS)},
  year      = {2024}
}

@inproceedings{ye2024chunkattention,
  title     = {{ChunkAttention}: Efficient Self-Attention with Prefix-Aware {KV} Cache and Two-Phase Partition},
  author    = {Ye, Lu and Tao, Ze and Huang, Yong and Li, Yang},
  booktitle = {Proceedings of the 62nd Annual Meeting of the Association for Computational Linguistics (ACL)},
  year      = {2024}
}

@article{cheng2025lmcache,
  title   = {{LMCache}: An Efficient {KV} Cache Layer for Enterprise-Scale {LLM} Inference},
  author  = {Cheng, Yihua and Liu, Yuhan and Yao, Jiayi and An, Yuwei and Chen, Xiaokun and Feng, Shaoting and Huang, Yuyang and Shen, Samuel and Du, Kuntai and Jiang, Junchen},
  journal = {arXiv preprint arXiv:2510.09665},
  year    = {2025}
}

@article{hu2024memserve,
  title   = {{MemServe}: Context Caching for Disaggregated {LLM} Serving with Elastic Memory Pool},
  author  = {Hu, Cunchen and Huang, Heyang and Hu, Junhao and Xu, Jiang and Chen, Xusheng and Xie, Tao and Wang, Chenxi and Wang, Sa and Bao, Yungang and Sun, Ninghui and Shan, Yizhou},
  journal = {arXiv preprint arXiv:2406.17565},
  year    = {2024}
}

@inproceedings{liu2024cachegen,
  title     = {{CacheGen}: {KV} Cache Compression and Streaming for Fast Large Language Model Serving},
  author    = {Liu, Yuhan and Li, Hanchen and Cheng, Yihua and Ray, Siddhant and Huang, Yuyang and Zhang, Qizheng and Du, Kuntai and Yao, Jiayi and Lu, Shan and Ananthanarayanan, Ganesh and Maire, Michael and Hoffmann, Henry and Holtzman, Ari and Jiang, Junchen},
  booktitle = {Proceedings of the ACM SIGCOMM Conference},
  year      = {2024}
}

@inproceedings{yao2025cacheblend,
  title     = {{CacheBlend}: Fast Large Language Model Serving for {RAG} with Cached Knowledge Fusion},
  author    = {Yao, Jiayi and Li, Hanchen and Liu, Yuhan and Ray, Siddhant and Cheng, Yihua and Zhang, Qizheng and Du, Kuntai and Lu, Shan and Jiang, Junchen},
  booktitle = {Proceedings of the Twentieth European Conference on Computer Systems (EuroSys '25)},
  year      = {2025}
}

@inproceedings{yu2022orca,
  title     = {Orca: A Distributed Serving System for Transformer-Based Generative Models},
  author    = {Yu, Gyeong-In and Jeong, Joo Seong and Kim, Geon-Woo and Kim, Soojeong and Chun, Byung-Gon},
  booktitle = {16th USENIX Symposium on Operating Systems Design and Implementation (OSDI '22)},
  year      = {2022}
}

@article{deepspeedfastgen2024,
  title   = {{DeepSpeed-FastGen}: High-throughput Text Generation for {LLMs} via {MII} and {DeepSpeed}-Inference},
  author  = {Holmes, Connor and Tanaka, Masahiro and Wyatt, Michael and Awan, Ammar Ahmad and Rasley, Jeff and Rajbhandari, Samyam and Aminabadi, Reza Yazdani and Qin, Heyang and Bakhtiari, Arash and Kurilenko, Lev and He, Yuxiong},
  journal = {arXiv preprint arXiv:2401.08671},
  year    = {2024}
}

\end{document}